\documentclass[12pt]{article}
\usepackage{graphicx}
\usepackage{amssymb}
\usepackage{cite}

\newcommand\pubnumber{}
\newcommand\pubdate{\today}

\def\iitm{Department of Physics\\
Indian Institute of Technology, Madras\\
Chennai 600036 \\
India}

\def\Title#1{\begin{center} {\Large #1 } \end{center}}
\def\Author#1{\begin{center}{ \sc #1} \end{center}}
\def\Address#1{\begin{center}{ \it #1} \end{center}}

\newcommand\pubblock{\rightline{\begin{tabular}{l} \pubnumber\\
         \pubdate  \end{tabular}}}
\newenvironment{Abstract}{\begin{quotation}  }{\end{quotation}}
\newenvironment{Presented}{\begin{quotation} \begin{center} 
             PRESENTED AT\end{center}\bigskip 
      \begin{center}\begin{large}}{\end{large}\end{center} \end{quotation}}
\def\Acknowledgements{\bigskip  \bigskip \begin{center} \begin{large}
             \bf ACKNOWLEDGEMENTS \end{large}\end{center}}

\def\beq{\begin{equation}}
\def\eeq#1{\label{#1}\end{equation}}
\def\eeqn{\end{equation}}

\def\beqa{\begin{eqnarray}}
\def\eeqa#1{\label{#1}\end{eqnarray}}
\def\eeqan{\end{eqnarray}}

\let\bar=\overbar

\def\Dslash{\not{\hbox{\kern-4pt $D$}}}
\def\dslash{\not{\hbox{\kern-2pt $\del$}}}

\def\msb{{\bar{\ssstyle M \kern -1pt S}}}

\begin{document}
\begin{titlepage}
\pubblock

\vfill
\Title{Direct $CP$ violation in hadronic $B$ decays}
\vfill
\Author{ Jim Libby}
\Address{\iitm}
\vfill
\begin{Abstract}
These proceedings review direct $CP$ violation in hadronic $B$ decays. The experimental results include those related to the measurement of the unitarity triangle angle $\gamma$ using $B\to D^{(*)}K^{(*)}$ decays and measurements of charmless hadronic $B$ decays. The results reported have been made by the BABAR, Belle and LHCb collaborations. Theoretical calculations related to these decays are also summarised. The importance of inputs from the charm sector in determining  $\gamma$ from $B\to D^{(*)}K^{(*)}$ decays is discussed.  In addition, the future prospects for these measurements will be reviewed. 
\end{Abstract}
\vfill
\begin{Presented}
The 8th International Workshop
on the CKM Unitarity Triangle (CKM 2014) Vienna, Austria, September 8-12, 2014
\end{Presented}
\vfill
\end{titlepage}
\def\thefootnote{\fnsymbol{footnote}}
\setcounter{footnote}{0}

\section{Introduction}
\label{sec:intro}
The study of direct $CP$ violation (DCPV) in hadronic $B$ decays can be divided into two distinct types of measurements.  One type are measurements of DCPV in $B^{-}\to DK^{-}$ \cite{CC} and related modes,\footnote{In $B^{-}$ decay the $D^{*}K^{-}$, $DK^{*-}$ and $D^{*}K^{*-}$ are sensitive to $\gamma$ as well. Only the hadronic part of the amplitude is different. In addition $B^{0}\to D^{(*)} K^{*0}(K^{+}\pi^{-})$ is also sensitve to $\gamma$ in a similar manner, with only the hadronic parameters being different.} which arise solely from the interference of first-order (tree) diagrams of differing weak and strong phase. Here, $D$ represents a $D^{0}$ or $\overline{D}^{0}$ decaying to the same final state. The tree-level nature of the amplitudes involved in  $B^{-}\to DK^{-}$ allows the theoretically clean extraction of  $\gamma$ (also denoted as $\phi_3$) $= -\arg{(V^{*}_{ub} V_{ud}/V^{*}_{cb} V_{cd})}$, an angle of the unitarity triangle, from the DCPV measurements. The other type are observations of DCPV in charmless hadronic $B$ decays, which arise from the interference of tree and higher-order (loop) amplitudes with differing phases.  Whereas DCPV measurements in charmless $B$ decay have sensitivity to non-Standard-Model physics due to the presence of loop diagrams. However, the presence of these loop diagrams also makes the  theoretical calculations much more complex. The distinct difference in the interfering amplitudes of $B^{-}\to DK^{-}$ and charmless decays motivates the separate discussions in Sections~\ref{sec:gamma} and \ref{sec:charmless}, respectively.  Recent measurements of these decays have been made by the $e^{+}e^{-}$ $B$ factories (BABAR and Belle) and LHCb. The prospects for future measurements with these experiments and the next generation -  Belle II and upgraded LHCb - are also discussed in Section~\ref{sec:outlook}. It should be noted that these proceedings focus upon measurements and theoretical advances made since the last CKM workshop held in 2012.
 
\section{DCPV in $B^{+}\to DK^{+}$}
\label{sec:gamma}
Improved knowledge of the unitarity triangle angle $\gamma$  is necessary for testing the Standard Model description of $CP$ violation.  The current precision on $\gamma$ is an order of magnitude worse than that on $\beta$ ~\cite{PDG} and it is the only measurement of the unitarity triangle that can be improved significantly by experimental advances alone. Sensitivity to $\gamma$ can be obtained by studying $CP$-violating observables in $B \to D K^{+}$ decays. There are two tree amplitudes contributing to $B^{-}\to DK^{-}$ decays: $B^{-}\to D^{0}K^{-}$ and $B^{-}\to \overline{D}^{0}K^{-}$. The amplitude for the second decay is both CKM and colour suppressed with respect to that for the first. The ratio of the suppressed to favoured amplitudes is written as 
\begin{displaymath}
 \frac{A(B^{-}\to \overline{D}^{0}K^{-})}{A(B^{-}\to D^{0}K^{-})} = r_{B} e^{i(\delta_B - \gamma)} \; ,
\end{displaymath}
where $r_B$ is the ratio of magnitudes and $\delta_B$ is the strong phase difference. The value of $r_B$ is approximately 0.1. The fact that the hadronic parameters $r_B$ and $\delta_B$ can be determined from data together with $\gamma$ makes these measurements essentially free of theoretical uncertainties. 

Several different types of $D$ decay are utilized to determine $\gamma$. Examples of $D$ decays include $CP$-eigenstates~\cite{GLW},  Cabibbo-favoured (CF) and doubly-Cabibbo-suppressed (DCS) decays~\cite{ADS},  self-conjugate modes \cite{GGSZ, BONDAR} and singly Cabibbo-suppressed (SCS) decays \cite{GLS}. The different methods are known by their proponents initials, which are given in Table~\ref{tab:naming}, along with the $D$ final states\footnote{$K^{0}_{S}\phi$ has also been included in early GLW measurements but has been dropped from more recent analyses given that the same data forms part of the $K^{0}_{S}K^{+}K^{-}$ sample, which can be studied with the GGSZ method.} that have so far been studied. 

\begin{table}[t]
\begin{center}
\begin{tabular}{l|cl}  
Type of $D$ decay &  Method name &  $D$ final states studied \\ \hline
$CP$-eigenstates & GLW & $CP$-even: $K^{+}K^{-}$, $\pi^{+}\pi^{-}$; $CP$-odd $K^{0}_{S}\pi^{0}$, $K^{0}_{S}\eta$   \\ 
$CF$ and $DCS$ & ADS & $K^{\pm}\pi^{\mp}$, $K^{\pm}\pi^{\mp}\pi^{0}$, $K^{\pm}\pi^{\mp}\pi^{+}\pi^{-}$ \\
Self-conjugate & GGSZ & $K^{0}_{S}\pi^{+}\pi^{-}$, $K^{0}_{S}K^{+}K^{-}$, $\pi^{+}\pi^{-}\pi^{0}$ \\
SCS & GLS & $K^{0}_{S}K^{\pm}\pi^{\mp}$ \\ \hline
\end{tabular}
\caption{Methods and $D$ decay modes used in $B^{-}\to DK^{-}$ measurements.} 
\label{tab:naming}
\end{center}
\end{table}

In the following four subsections (i) advances in understanding the theoretical cleanliness of these modes to extract $\gamma$, (ii) recent results, (iii) external inputs and (iv) the world average, are reviewed in turn.   
 
\subsection{The ultimate precision}
\label{SEC:ULTIMATE}
Significant corrections to the value of $\gamma$ extracted from $B^{-}\to DK^{-}$ might arise from two sources: mixing and DCPV in $D$ decay and higher-order diagrams that contribute with differing 
CKM matrix elements to the tree diagrams. Several studies of the impact of mixing and DCPV in charm decays have been made \cite{MIXING_GAMMA_1,MIXING_GAMMA_2,MIXING_GAMMA_3,MIXING_GAMMA_4,CHARM_DCPV_1, CHARM_DCPV_2, CHARM_DCPV_3, CHARM_DCPV_4, MIXING_GAMMA_5}. These studies show that  $\gamma$ can be extracted without bias as long as appropriate modifications of the formalism are made and the measured values of the mixing and DCPV parameters are included as external inputs. Even if the effect of mixing is neglected the size of the induced bias is less than $1^{\circ}$ \cite{MIXING_GAMMA_5}. 

 Measurements of $\gamma$ can be made using the $B^{-}\to D\pi^{-}$ decay mode, which has sensitivity to $\gamma$ in the same manner as $B^{-}\to D K^{-}$.  However, the size of the DCPV is much smaller due to the ratio of the suppressed to favoured amplitudes being approximately 0.005. The reduced sensitivity due to the smaller interference is somewhat compensated by the much larger branching fraction for $B^{-}\to D\pi^{-}$ compared to $B^{-}\to DK^{-}$ \cite{PDG}. However, $D$ mixing and DCPV must be accounted for carefully in $B^{-}\to D\pi^{-}$ measurements of $\gamma$ because the bias on the extracted value of $\gamma$ would be $\mathcal{O}(10^{\circ})$ otherwise \cite{MIXING_GAMMA_5}.

 The impact of the irreducible uncertainty due to higher-order diagrams has been studied recently \cite{BRODZUPAN} to ascertain the ultimate precision with which $\gamma$ can be measured. Second-order weak-box diagrams are the first processes to have a differing CKM dependence from the tree diagrams. An effective-field-theory calculation of the shift in $\gamma$, $\delta\gamma$, including resumming the large logarithms of $m_b/m_W$ in the corrections to the Wilson coefficients, gives $\delta\gamma\sim 2\times 10^{-8}$. Long distance contributions are at most a factor of a few larger than the calculated short-distance contribution. Therefore, the relative shift in $\gamma$ due to the neglect of these weak-box diagrams is $\lesssim 10^{-7}$, which is many orders of magnitude below the experimental precision anticipated at future experiments. Since the workshop the ultimate theoretical precision due to electroweak effects for $B\to D\pi$ decays has been investigated further \cite{BROD}. Due to cancellations in these corrections the relative shift in $\gamma$ from $B\to D\pi$ may be enhanced compared to $B\to DK$ up to $10^{-4}$. 
   
The effect of new physics in tree-level amplitudes has also been reported recently \cite{GAMMANP}. Accounting for current experimental bounds, a new-physics induced shift of up to $4^{\circ}$ on the Standard Model value of $\gamma$ is still possible. This result is a strong motivation for the $1^{\circ}$ precision being pursued by the future experimental programme discussed in Sec.~\ref{sec:outlook}. 

\subsection{Review of recent $B\to D^{(*)}K^{(*)}$ measurements}
\label{SEC:GAMMARESULTS} 
There is only a single new measurement related to $\gamma$ from the $e^{+}e^{-}$ $B$ factories since the last CKM workshop. An ADS analysis of $B^{-}\to D(K^{\pm}\pi^{\mp}\pi^{0})K^{-}$ has been reported by Belle \cite{MINAKSHI} that uses the full data set of $772\times 10^6$ $B\bar{B}$ pairs. The first evidence of the suppressed decay $B^{-}\to D(K^{+}\pi^{-}\pi^{0})K^{-}$ is reported with a significance of 3.2 standard deviations. No statistically significant DCPV is observed but given the yield is of a similar magnitude to the $B^{-}\to D(K^{+}\pi^{-})K^{-}$ decay it is a promising mode for Belle II in particular. Also, the interpretation of the measurements in terms of $r_B$, $\delta_B$ and $\gamma$ requires information about the strong dynamics of the $D$ decay. The relevant $D$ decay parameters, the coherence factor $R$ and average strong-phase difference $\delta_D$ \cite{ATWOODSONI}, have been measured in quantum correlated  $D^{0}\overline{D}^{0}$ production by CLEO-c \cite{WINGS,SOL}. The large value of 
$R=0.82\pm 0.07$ for this decay means that the sensitivity to $\gamma$ is not diluted significantly in this channel even though it is a multibody $D$ decay. This measurement has been included in a combination of all 
Belle measurements related to $\gamma$ and the external charm physics inputs have been updated since the last average was reported \cite{KARIMCKM2012}. The value of $\gamma$ from Belle alone is $(73^{+13}_{-15})^{\circ}$ and is dominated by the GGSZ measurement of $B^{-}\to D^{(*)}(K^{0}_{S}\pi^{+}\pi^{-})K^{-}$ \cite{BELLEGGSZ}. 

LHCb have reported several new and updated measurements related to $\gamma$ since the last CKM workshop. The first measurement of the GLS mode $B^{\pm}\to D(K^{0}_{S}K^{\pm}\pi^{\mp})K^{\pm}$ has been reported \cite{LHCBGLS} based on a data sample corresponding to an integrated luminosity of 3~fb$^{-1}$. The measurement is independent of any amplitude model and the $\gamma$-sensitive observables can be interpreted using $R$ and $\delta_D$ measurements reported by CLEO-c \cite{CLEOK0SKPI}. Furthermore, results are reported solely for events in which the $K^{0}_{S}\pi^{\mp}$ invariant mass is consistent with that of the $K^{*}(892)^{+}$; this region is more sensitive to DCPV because it has a larger value of $R$. No statistically significant DCPV is observed. 

A new GLW and ADS measurement of $\bar{B}^{0}\to D K^{*0}(K^{-}\pi^{+})$ decays, where the $D$ decays to $\pi^{+}\pi^{-}$, $K^{+}K^{-}$ or $K^{\pm}\pi^{\mp}$, has been reported by LHCb \cite{LHCBB0TODKST}. The full 3~fb$^{-1}$ data sample is used. These self-tagged decays of the neutral $B^{0}$ are of particular interest because the interfering CKM favoured and CKM suppressed amplitudes are both colour suppressed. Therefore, the magnitude of the ratio of the suppressed to favoured amplitudes $r_B(DK^{*})$ is expected to be enhanced compared to $r_B(DK)$, resulting in larger DCPV effects. However, with both amplitudes colour suppressed the branching fraction is substantially smaller than that for $B\to DK$. The analysis provides the single most precise measurement of $r_B(DK^{*})=0.240^{+0.055}_{-0.048}$, but no evidence for DCPV is observed.\footnote{LHCb have also reported the first measurement of $\gamma$ in $B_{s}^{0}\to D_s^{\pm}K^{\mp}$ decays \cite{LHCBBSTODSK}. However, as the sensitivity to $\gamma$ comes via the interference of mixing and decay amplitudes and requires time-dependent techniques the mode is outside the remit of this working group. 
The reported value of $\gamma = (115^{+28}_{-43})^{\circ}$ is included in the combined value of $\gamma$ presented by LHCb but it has limited weight in the combination due to the large uncertainties compared to the $B\to DK$ measurements.}
   
The most important recent contribution to the combined measurement of $\gamma$ is the updated GGSZ measurement of $B^{+}\to D(K^0_S h^{+}h^{-})$, where $h$ is a pion or a kaon, to the full $3~\mathrm{fb}^{-1}$ LHCb data set  \cite{LHCBGGSZUPDATE}. A model-independent technique to determine $\gamma$ is used. To achieve model-independence the data 
must be binned in the $K^0_S h^{+}h^{-}$ Dalitz plot and combined with measurements of the strong-phase difference within these bins as reported by the CLEO Collaboration \cite{CLEOC_CISI}. Despite the small loss of statistical sensitivity arising from a binning the data, which is not required for the model-dependent method, there is a significant systematic advantage to the model-independent approach. The systematic uncertainty related to the ansatz of the amplitude model in the model-dependent approach is replaced by an uncertainty related to the strong-phase measurements in the model-independent method.  The model-related systematic uncertainty would ultimately limit the precision with which $\gamma$ can be determined in this mode, whereas the CLEO-c uncertainty is largely statistical and can be improved as described in Sec.~\ref{SEC:SHITTINGDOG}. The result of the measurement is $(\gamma = 62^{+15}_{-14})^{\circ}$ and $r_B = 0.080^{+0.19}_{-0.21}$.  

\begin{figure}[htb]
\centering
\includegraphics[height=3.0in]{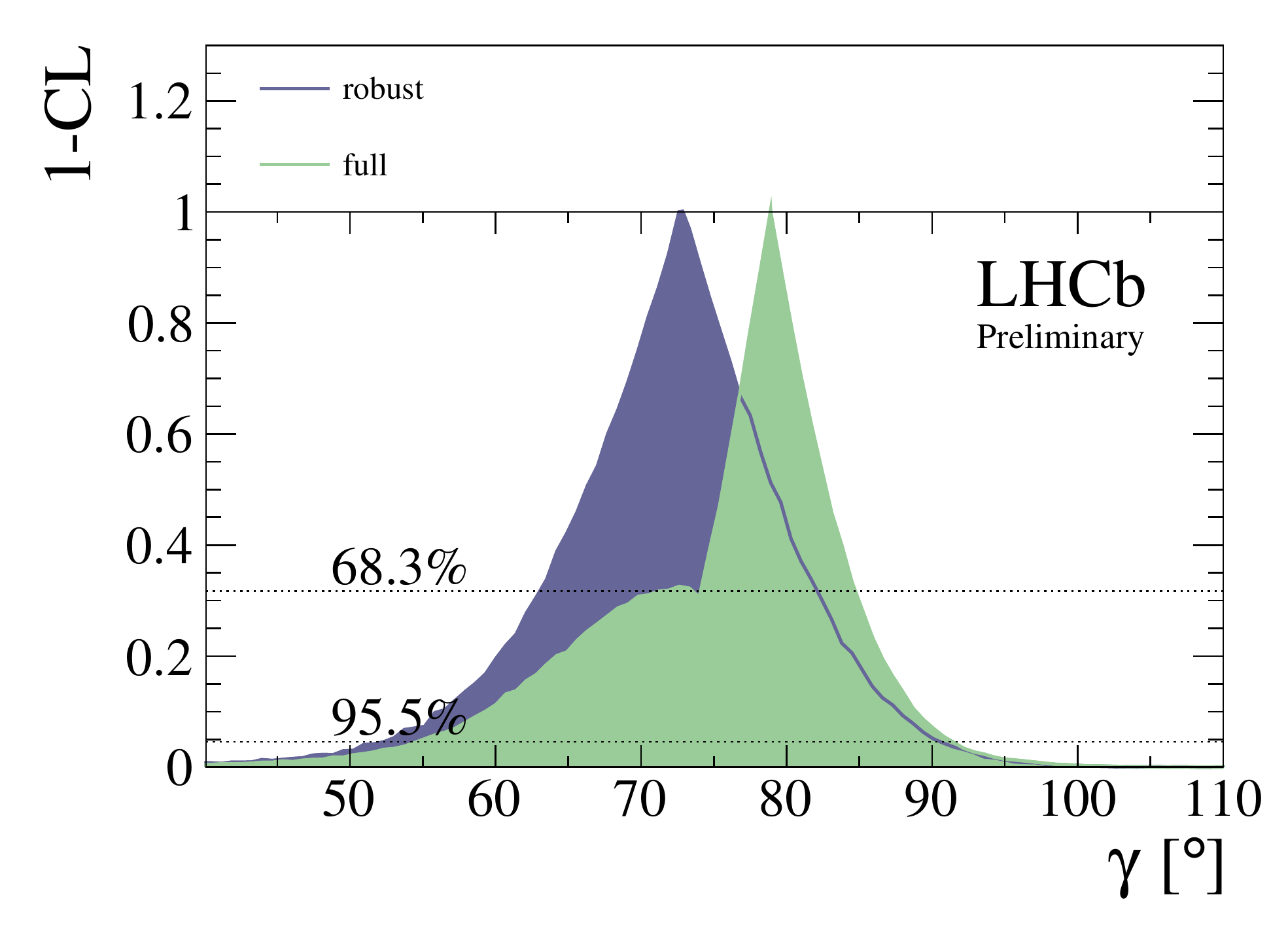}
\caption{$1-CL$ as a function $\gamma$ for the combination of LHCb results. Robust (blue) refers to the combination of $B\to DK$ measurements alone while the full (green) includes $B\to D\pi$ measurements as well.}
\label{fig:lhcb_combo}
\end{figure}

LHCb reported a new combination of measurements to determine $\gamma$ \cite{KARBACH}. Aside from the recent new and updated measurements discussed in this section, the ADS and GLW measurements made with 1~fb$^{-1}$ of data \cite{LHCBADS} are important contributions to the combination. In addition, several external inputs related to charm decay - mixing parameters, $CP$ asymmetries, strong-phase parameters and branching fractions - are also updated in the combination. (See Ref. \cite{KARBACH} for the details.) The combination is performed with (full) or without (robust) the ADS and GLW measurements of $B^{-}\to D \pi^{-}$ \cite{LHCBADS}.  The robust combination gives $\gamma = (72.9^{+9.2}_{-9.9})^{\circ}$, the most precise determination from a single experiment. 
The full combination favours a second solution at $78.9^{\circ}$ with a smaller $68\%$ confidence level (CL) than the robust determination, but the $95\%$ CL is almost identical for the full and robust combinations as shown in Fig.~\ref{fig:lhcb_combo}. The sensitivity of the combination to the  $B^{-}\to D \pi^{-}$ measurements is related to the large prefered value of $r_B(D\pi)$ of 0.027, which is approximately five times larger than the expectation. However, there is a second solution close to the expectation and the $95\%$ CL is [0.001,0.040], which means more data is required to resolve the true value. The update of the ADS and GLW analyses to the $3~\mathrm{fb}^{-1}$ data set is greatly anticipated. 
 
\subsection{Auxiliary measurements}
\label{SEC:SHITTINGDOG}
The precise determination of $\gamma$ using $B^{-}\to DK^{-}$ is reliant upon external inputs from the charm sector. The accurate of determination of charm-mixing parameters  \cite{GERSABECK} means that any bias from this source in the determination of $\gamma$ can be corrected for as discussed in Sec.~\ref{SEC:ULTIMATE}. In addition, $D$ meson branching fractions of both CF and DCS decays provide important inputs to ADS measurements \cite{DHAD,ERIC}. 

 However, the most important auxiliary measurements are related to $D$ decay strong-phases, which are an essential input to interpret the measurements related to $\gamma$. In principle these parameters could be extracted from the $B$ data along with $\gamma$, $\delta_B$ and $r_B$, but the sensitivity to $\gamma$ would be diluted significantly. Therefore, measurements of the strong-phases are taken from elsewhere. The strong-phase difference between the $D^{0}$ and $\overline{D}^{0}$ decays to $K^{+}\pi^{-}$ is required for the two-body ADS measurement and it is accurately determined using the combination of charm-mixing measurements \cite{GERSABECK}. For multibody ADS measurements two parameters must be determined due to the variation of the strong-phase difference over the allowed phase-space: the coherence factor $R$ and average strong-phase difference $\delta_D$. Recently there has been a new analysis to determine the $R$ and $\delta_D$ for $D\to K^{-}\pi^{+}\pi^{0}$ and $D\to K^{-}\pi^{+}\pi^{+}\pi^{-}$ \cite{SOL}, which uses quantum-correlated $D^{0}\overline{D}^{0}$ pairs produced at the $\psi(3770)$. (For a comprehensive review of quantum-correlated measurements relevant to $\gamma$ see Ref.~\cite{BRIERE}.) At the $\psi(3770)$ the $D$ decay of interest is tagged in events where the other $D$ decays to a $CP$-eigenstate, a state with a kaon of opposite or same-sign charge as the signal or $K^{0}_{S,L}\pi^{+}\pi^{-}$. The last of these tags is an addition since the first determination of $R$ and $\delta_D$ reported by the CLEO-c collaboration \cite{LOWREY}. The updated results are used to perform the combinations reported elsewhere in these proceedings.

The model-independent GGSZ method requires two parameters related to the strong-phase difference to be determined for each bin of the Dalitz plot. Such measurements have been reported by the CLEO Collaboration \cite{CLEOC_CISI} using a data sample corresponding to an integrated luminosity of  $818~\mathrm{pb}^{-1}$. These measurements have been used by both the Belle~\cite{BELLEMODIND} and LHCb~\cite{LHCBGGSZUPDATE} collaborations to determine $\gamma$ from $B^{-}\to DK^{-}$ data. The 
systematic uncertainty on $\gamma$ related to the statistical precision of the CLEO measurements is not dominant at present, but will become much more significant with the future running of LHCb and Belle II. Therefore, improvements in the measurements of the strong phase parameters are desirable. BESIII has accumulated an integrated luminosity of $2.92~\mathrm{fb}^{-1}$ at the $\psi(3770)$ which is 
3.5 times larger than that analysed by CLEO. Preliminary results for the $D\to K^{0}_{S}\pi^{+}\pi^{-}$ parameters using the same binning as CLEO have been reported \cite{AMBROSE}, which give a significant improvement in the statistical uncertainty on the measurements. BESIII can accumulate around $4~\mathrm{fb}^{-1}$ of integrated luminosity per year of running at the $\psi(3770)$; therefore, a two year run at the $\psi(3770)$ by BESIII would reduce the uncertainty on $\gamma$ from the determination of strong phases in the GGSZ method to a negligible level.

Quantum-correlated measurements are also opening up new pathways to determining $\gamma$. A measurement of the $CP$ content of $D\to \pi^{+}\pi^{-}\pi^{0}$ and $D\to K^{+}K^{-}\pi^{0}$ \cite{HHPI0} using the full CLEO-c $\psi(3770)$ data set has shown that $D\to \pi^{+}\pi^{-}\pi^{0}$ is $(96.8\pm 1.7 \pm 0.6)\%$ $CP$-even. Therefore, this mode can be used as an additional GLW measurement to augment $D\to h^{+}h^{-}$, given it has a significantly larger branching fraction \cite{PDG}. 
   
\subsection{World average}
\begin{figure}[htb]
\centering
\begin{tabular}{cc} 
\includegraphics[width=0.49\columnwidth]{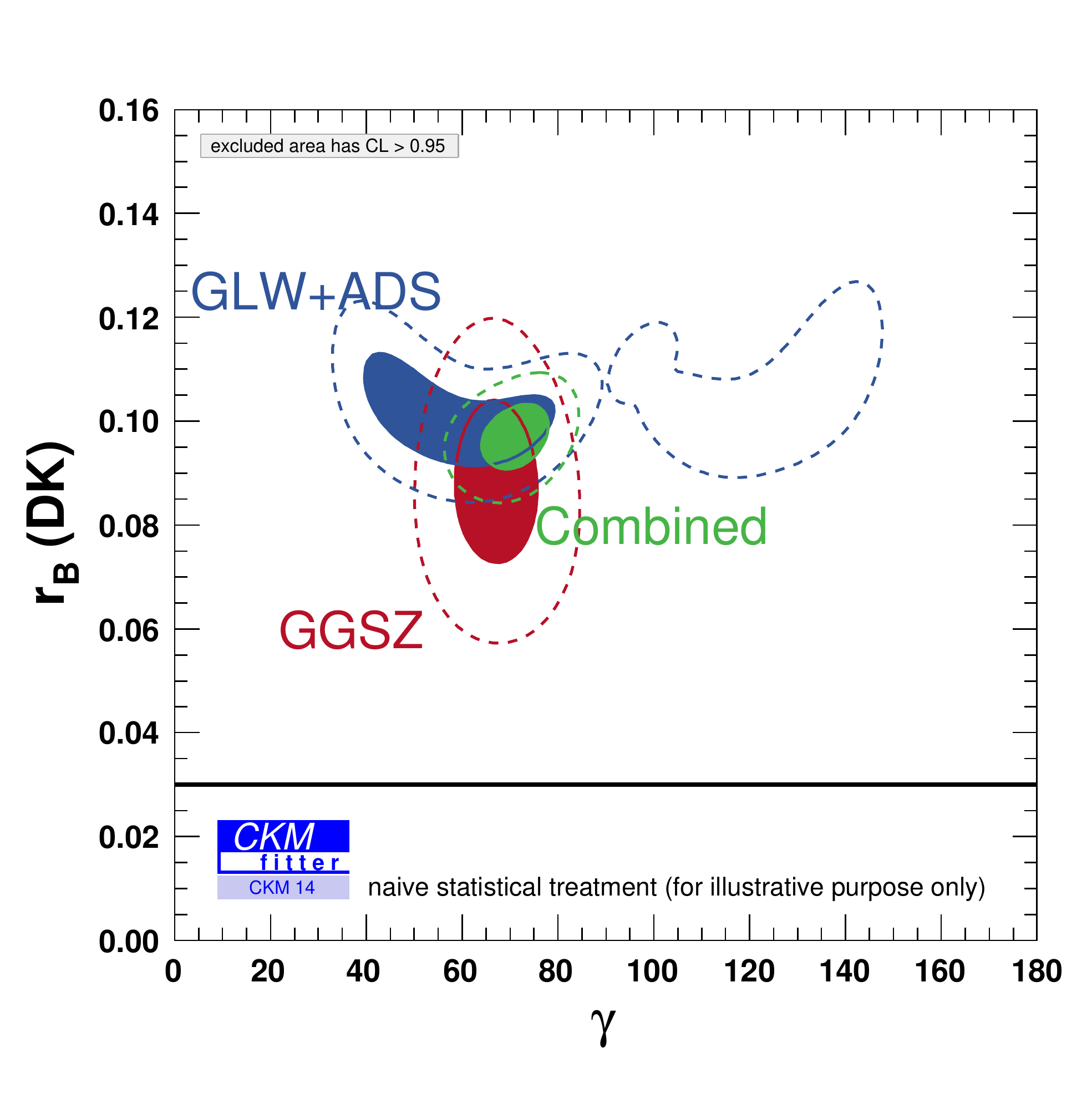} & 
\includegraphics[width=0.49\columnwidth]{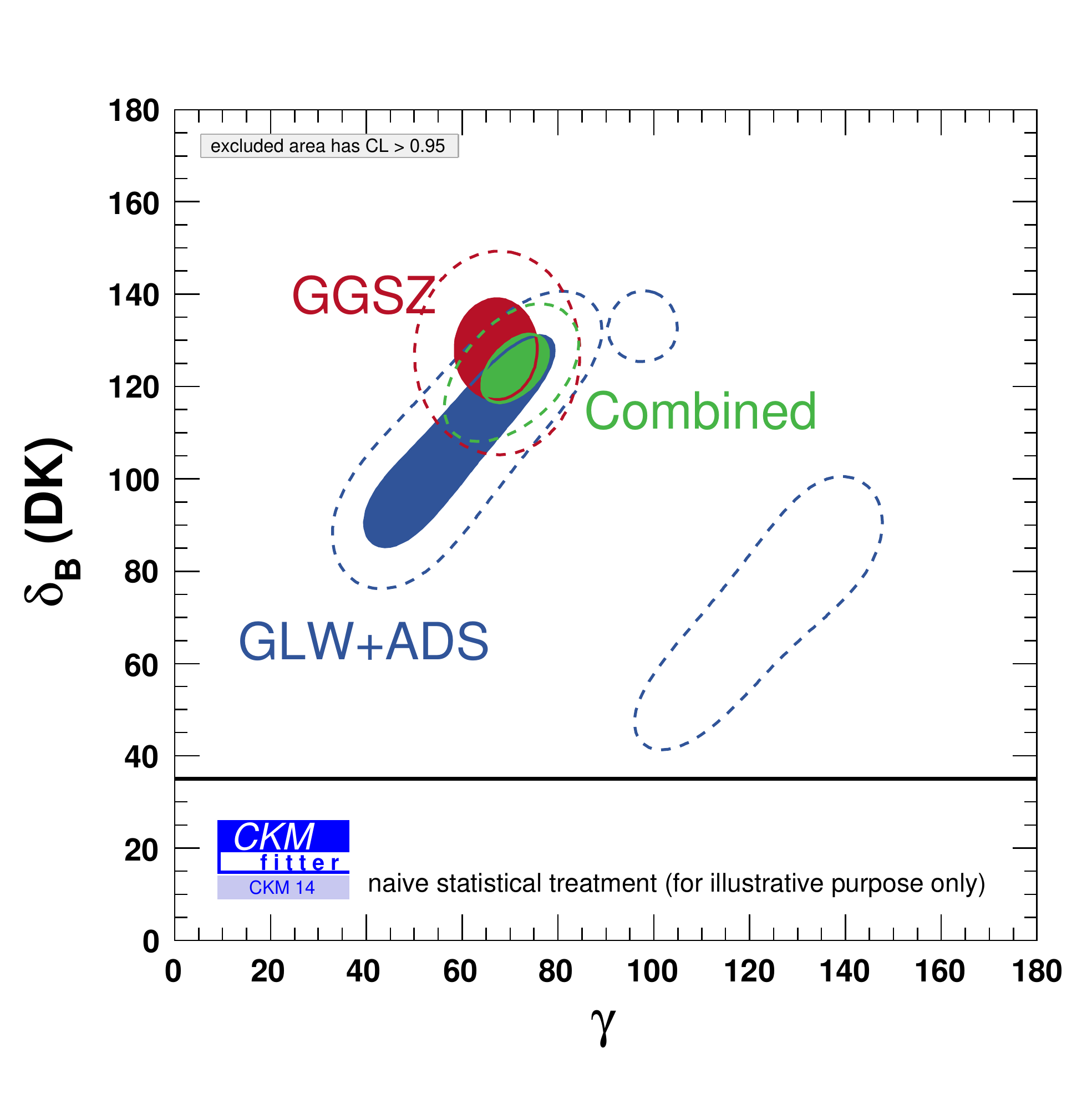}
\end{tabular}
\caption{The (solid) $1\sigma$ and (dashed) 95\% CL for (blue) GLW and ADS measurements, (red) GGSZ and (green) combined. The (left) $\gamma~vs.~r_B$ and (right) $\gamma~vs.~\delta_B$ parameter spaces are shown.}
\label{FIG:KARIM}
\end{figure}

Most of the measurements related to $\gamma$ descussed in Sec.~\ref{SEC:GAMMARESULTS} from the Belle and LHCb collaborations have been combined with those from BABAR \cite{BABARGAMMA}  along with the auxiliary inputs to form a new world average \cite{CKMFITTER}. The $B\to D\pi^{-}$ and GLS results from LHCb have not been included. The different constraints on $\gamma$, $r_B$ and $\delta_B$ derived from the ADS/GLW and GGSZ methods are shown in Fig.~\ref{FIG:KARIM}. The world averages for are:
\begin{eqnarray*}
  \gamma & = & \left(73.2^{+6.3}_{-7.0}\right)^\circ \; ,\\
  r_B & = & 0.0970^{+0.0062}_{-0.0063} \; \mathrm{and} \\
 \delta_B & = & \left(125.4^{+7.0}_{-7.8}\right)^\circ \; .\\
\end{eqnarray*} 

\section{DCPV in charmless $B$ decay}
\label{sec:charmless}
Charmless $B$ decays have a rich phenomenology with a plethora of final states to study. The dominant amplitudes contributing to the decays differ mode-to-mode, as do the nature of the final states in terms of the multiplicity and spin of the hadrons involved. In principle there is great potential to observe high-mass new physics contributing to the penguin amplitudes that mediate these decays. However, robust theoretical calculations of the Standard Model expectations are required to realise this potential. Therefore, we will review the theoretical advances, before summarising recent measurements of hadronic charmless $B$ decay by the Babar, Belle and LHCb Collaborations. 
\subsection{Theoretical advances}
\label{SEC:CHARMLESSTHEORY}
There are two different approaches to calculating the rates and asymmetries in charmless $B$ decays: flavour symmetries, such as U-spin, and heavy-quark expansions, such as QCD factorisation \cite{QCDF}. The next-to-next-to-leading-order (NNLO) QCDF calculation  of corrections to observables related to the $B\to PP$, $B\to PV$ and $B\to VV$ decays has been largely completed \cite{NNLO1,NNLO2,NNLO3,NNLO4,NNLO5,NNLO6,NNLO7,NNLO8}, apart from two loop corrections to the hard-scattering kernals in penguin amplitudes. Here $P$ and $V$ represent a pseudoscalar and vector meson, respectively. These calculations have allowed the first predictions at NNLO of observables dominated by trees \cite{NNLO8, NNLO9}. Recently preliminary results have been released for the missing two-loop corrections \cite{NNLO10}. The corrections are significant; for example the charmed-penguin amplitude for $B\to\pi\pi$ has 30\% and 80\% corrections to the real and imaginary parts, respectively. The updated predictions of observables with this final piece of the NNLO calculation are eagerly anticipated.

\subsection{Recent experimental results of charmless $B$ decay}
\label{SEC:CHARMLESSEXP}
 In these proceedings only the most significant measurements of charmless $B$ decays presented at the workshop will be summarised. Results on other modes were also presented by the Belle \cite{BELLEETAPRIMEKST,BELLEKSTKST} and BABAR \cite{BABAROMEGAOMEGA} and LHCb \cite{LHCBPHIKST} Collaborations. 

 The most significant new measurement of $B\to PP$ decays is a preliminary result for the  $B^{0}\to \pi^{0}\pi^{0}$ branching fraction from the Belle Collaboration that exploits the full $\Upsilon(4S)$ data set. This result can be used, in combination with other measurements of the branching fractions and asymmetries for  $B\to\pi\pi$ decays, to determine $\alpha$ \cite{GLPIPI}. The measured branching fraction is $(0.90\pm 0.12 \pm 0.10) \times 10^{-6}$ which is significantly lower than the previous Belle measurement \cite{BELLEPI0PI0}. The reason for the change is the discovery of a peaking background related to pile-up of a electromagnetic deposits from  earlier Bhabha events with a $B\overline{B}$ event in the old analysis; this is now excluded by including timing information for the electromagnetic calorimeter hits. In addition, the measured value is in agreement with the theoretical predications discussed in Sec~\ref{SEC:CHARMLESSTHEORY}. However, the measurement disagrees with that from the BABAR collaboration \cite{BABARPI0PI0} by 3.2 standard deviations; this discrepancy will only be resolved by an analysis at Belle II with a significantly larger data set.

\begin{figure}[htb]
\centering
\includegraphics[width=0.95\columnwidth]{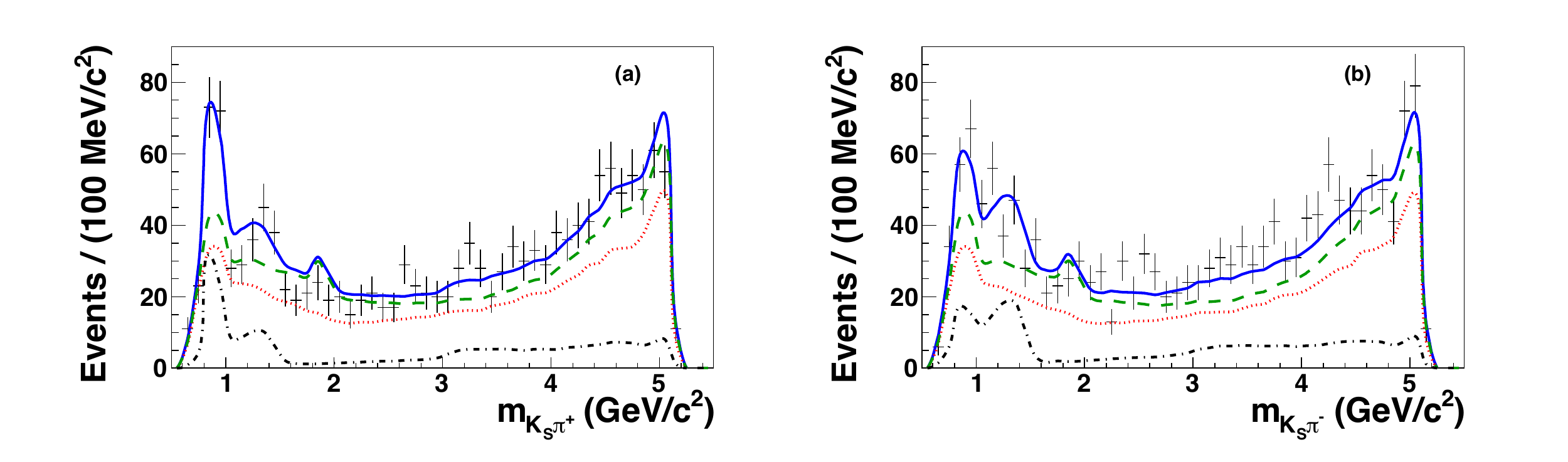} 
\caption{Amplitude fit projection on the $K^{0}_{S}\pi^{\pm}$ invariant mass distribution for (a) $B^{+}\to K^{0}_{S}\pi^{+}\pi^{0}$ and (b) $B^{-}\to K^{0}_{S}\pi^{-}\pi^{0}$ decays. The different probability density function components are: signal (black dot-dashed line), continuum background (red dotted line), total background (green dashed line) and total (blue solid line).}
\label{FIG:ELI} 
\end{figure}

The significant difference in the world average values of the DCPV asymmetries in $B^{+}\to K^+\pi^{0}$ and $B^{0}\to K^{+}\pi^{-}$ \cite{HFAG}, which disagrees with the theoretical prediction of zero, is referred to as the ``$K\pi$ puzzle''. A way to study this conundrum further is to measure the equivalent asymmetries in the $B\to K^{*}(892)\pi$ decays \cite{KSTPITHEORY}. The BABAR collaboration presented the first study of DCPV in $B^+\to K^{0}_{S}\pi^{+} \pi^{0}$ decays, which includes a determination of the $CP$ asymmetry in $B^{+}\to K^{*+}(892)\pi^0$ decays. The analysis uses the full $\Upsilon(4S)$ data set, corresponding to an integrated luminosity of $429~\mathrm{fb}^{-1}$. The selected data sample contains approximately 1000 signal events that are fit to an isobar amplitude model with five components. The direct $CP$ asymmetry, $\mathcal{A}_{CP}$,  is determined for the different components and it is found that 
\begin{displaymath}
\mathcal{A}_{CP}(B^{+}\to K^{*+}(892)\pi^0) = -0.52\pm 0.14 \pm 0.04 \pm 0.04 \;,
\end{displaymath}
which differs from zero by 3.4 standard deviations. Here the third uncertainty is related to the ansatz of the isobar model. The asymmetry can be seen clearly in the fit projection onto the distribution of the invariant mass of $K^{0}_{S}\pi^{\pm}$ shown in Fig.~\ref{FIG:ELI}.
In addition, the value of  
$\mathcal{A}_{CP}(B^{+}\to K^{*0}(892)\pi^{+})$ found is consistent with zero, which is the Standard Model prediction. Including this result the value of $\Delta\mathcal{A}_{CP}(K^{*}\pi) = \mathcal{A}_{CP}(K^{*+}\pi^{0})-\mathcal{A}_{CP}(K^{*+}\pi^{-})= -0.16\pm 0.14$ \cite{BEACHLATHAM}, which is consistent with the Standard Model prediction.
 
\begin{figure}[htb]
\begin{centering}
\begin{tabular}{cc} 
\includegraphics[width=0.49\columnwidth]{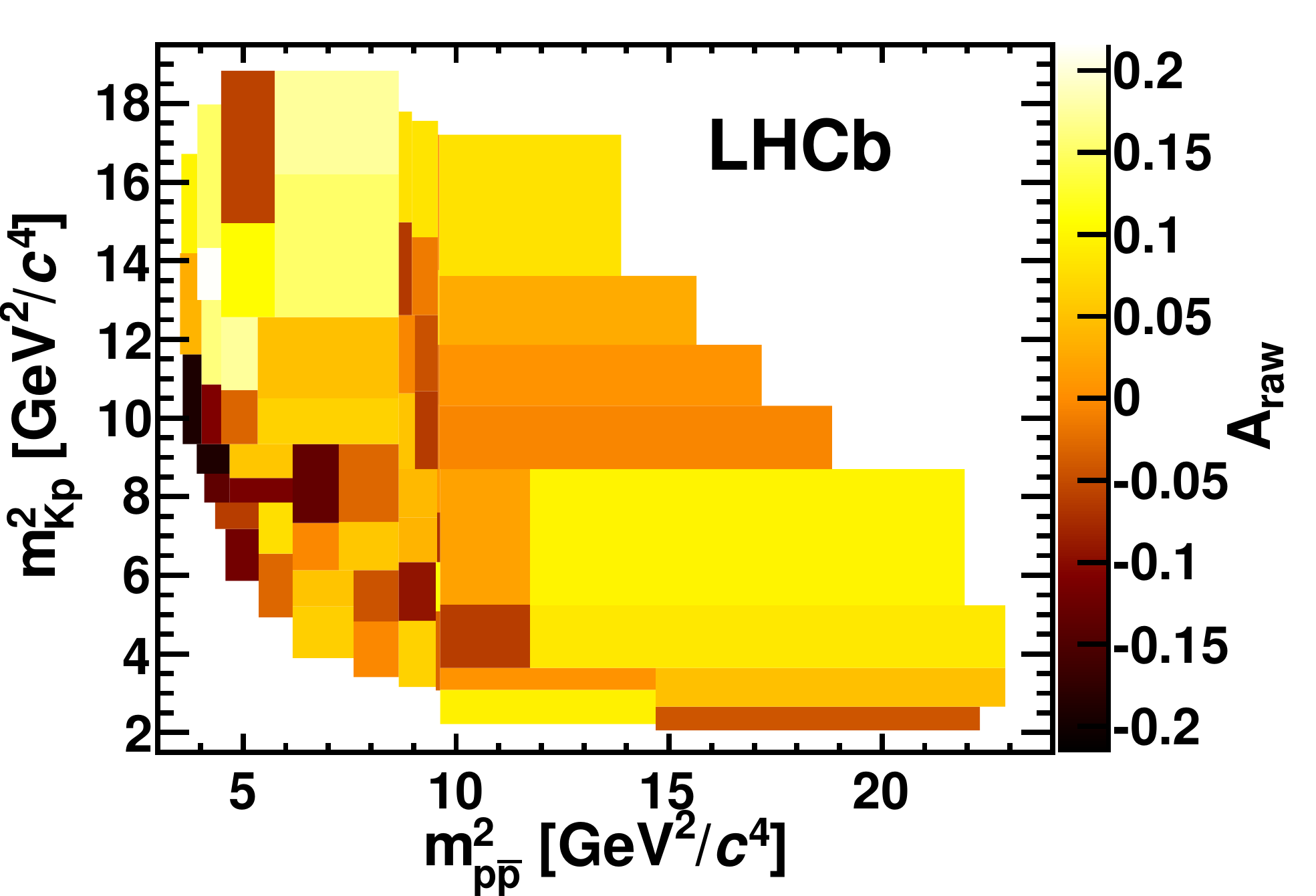} & 
\includegraphics[width=0.49\columnwidth]{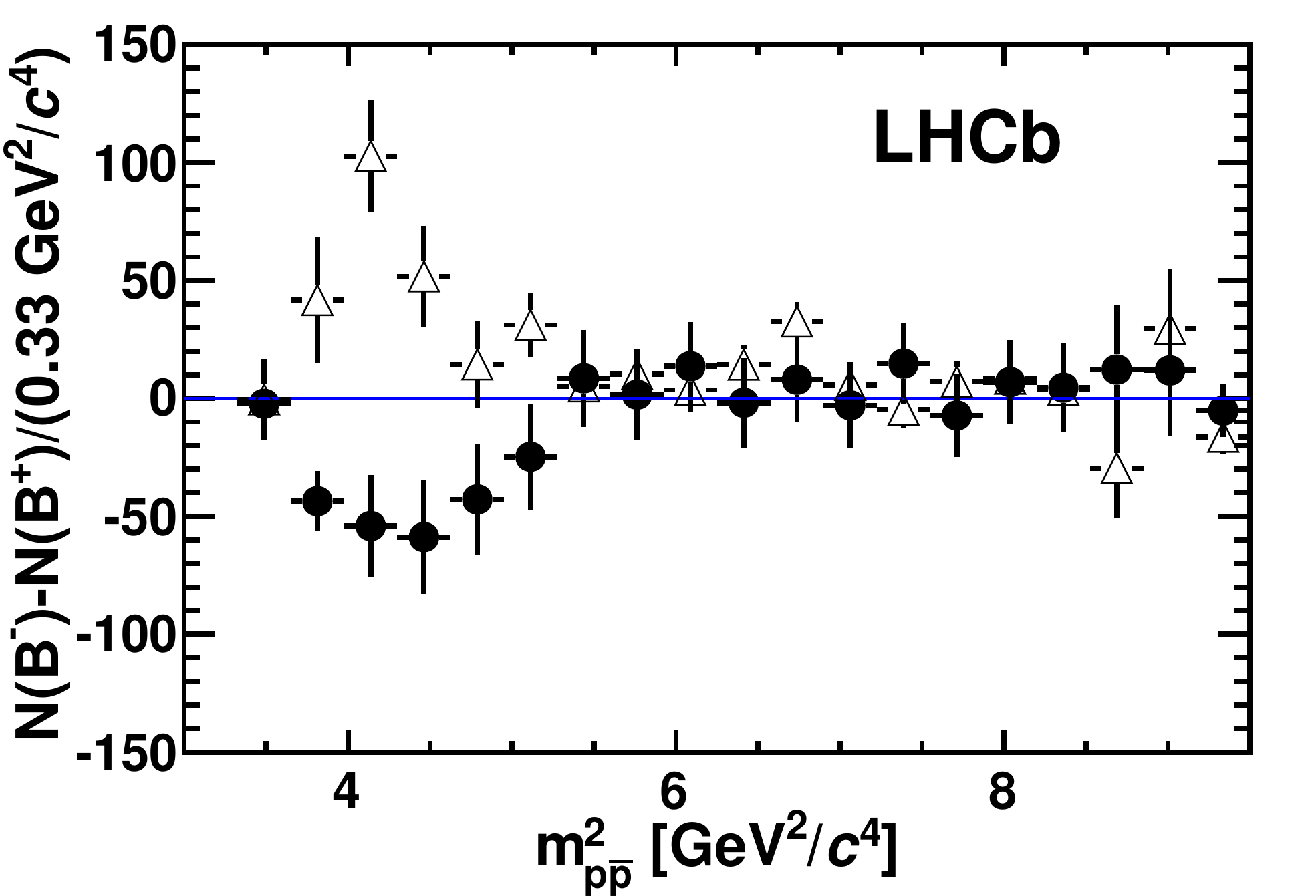}
\end{tabular}
\end{centering}
\caption{Raw asymmetry of  $B^{+}\to p\bar{p} K^{+}$ decays  (left) over the whole Dalitz plot and (right) $N(B^{-}) -N(B^{+})$ as  a function of $m_{p\bar{p}}^2$ for $m_{Kp}^2 < 10~\mathrm{GeV}^2/c^4$ (black filled circles) and $m_{Kp}^2 < 10~\mathrm{GeV}^2/c^4$ (open triangles).}
\label{FIG:MARC}
\end{figure}

The LHCb collaboration have updated their observations of DCPV in $B^{0}\to K^{+}\pi^{-}$, $B^{0}_{s}\to K^{-}\pi^{+}$ \cite{LHCBKPI} and $B^+\to h^+ h^- h^+$ \cite{LHCBHHH}, where $h$ is a $\pi$ or $K$, since the last CKM workshop using larger datasets. They have also studied DCPV in baryonic decays of a $B$ meson \cite{LHCBPPBARH}. Using the excellent particle identification capabilities of the detector, combined with a multivariate discriminant to suppress combinatoric background, clean signal samples of $B^{+}\to p \bar{p} K^{+}$ and $B^{+}\to p \bar{p} \pi^{+}$ of approximately 19,000 and 2000 events, respectively, are selected. The background is subtracted using an event-by-event reweighting \cite{SPLOT} so that the Dalitz plot can be used to determine branching fractions for intermediate states and search for 
DCPV. The raw asymmetry for $B^{+}\to p \bar{p} K^{+}$ in bins over the Dalitz plot is shown in Fig~\ref{FIG:MARC}; significant deviations from zero can be seen in several bins. The asymmetries can be seen more clearer in the plots showing the projection of the number of $B^{+}$ candidates, $N(B^{+})$, subtracted from the number of $B^{-}$ candidates, $N(B^{-})$,  as a function of the $p\bar{p}$ invariant-mass squared $m_{p\bar{p}}^2$ in two different regions of $pK$ invariant-mass squared. The asymmetry in the region with $m_{p\bar{p}}< 2.85~\mathrm{GeV}/c^2$ and $m_{Kp}^2>10~\mathrm{GeV}^2/c^4$ is $(9.6\pm 2.4 \pm 0.4)\%$, which 4.2 standard deviations away from zero; this is the first observation of DCPV in a baryonic decay of a $B$ meson.

\section{Outlook and conclusions}
\label{sec:outlook}
Significant progress has been made since the last CKM workshop both experimentally and theoretically. 
The world average for $\gamma$ is around $6^{\circ}$ and will improve further once all measurements from the LHCb $3~\mathrm{fb}^{-1}$ dataset have been completed. In relation to charmless decays the completion of NNLO calculations for $B\to PP$ is imminent and the updates to the predictions for branching fractions and asymmetries will be an important step in establishing whether new physics is lurking in these modes. Experimentally, the final word from the $e^{+}e^{-}$ $B$ factories  and new results from LHCb are establishing DCPV in new final states, such as those to baryons.

However, there is still much to be done to better understand DCPV in hadronic $B$ decays. The desire is to produce a determination of $\gamma$ with a precision similar to that on $\beta$, which is of the order of $1^{\circ}$. Even with the statistics accumulated by LHCb during the LHC13 run, this target will not be achieved. Therefore, the next generation of experiments - LHCb upgrade\cite{LHCBUPGRADE} and Belle II \cite{BELLEII} - will be the facilities that produce measurements of this precision. The LHCb upgrade, which will start taking data  in 2020, includes changes in the first-level trigger architecture such that the whole detector can be readout at the collision rate of 30 MHz and all trigger algorithms are performed in software. This change is advantageous for hadronic modes in particular as they will no longer saturate the bandwidth at lowest level of the trigger. It is anticipated that with a sample corresponding to an integrated luminosity of 50~fb$^{-1}$ an upgraded LHCb will be able to measure $\gamma$ with a precision of around $1^{\circ}$.

The Belle II detector will operate from 2017 at the upgraded SuperKEKB with the goal of collecting a 50~ab$^{-1}$ data set by 2023. A principal feature of the upgraded Belle II detector for measurements of DCPV is the enhanced particle identification capability, via a quartz based time-of-propagation Cherenkov detector and a forward  aerogel ring-imaging 
Cherenkov detector; this improvement will halve the $\pi\to K$ misidentification rate while increasing the kaon identification efficiency from 88\% to 94\% compared to Belle. In addition upgraded readout electronics for the electromagnetic calorimeter will improve the $\pi^{0}$ energy resolution, which will be helpful for many charmless studies in particular. The projected Belle II data set should allow for a measurement of $\gamma$ with a precision of at least $1.5^{\circ}$. Also, significant light can be shed on the ``$K\pi$ puzzle" with this data set. The sum rule introduced in Ref. \cite{KPISUMRULE} is a theoretically robust test of the consistency of the $B\to K\pi$ branching fractions and asymmetries. The Belle Collaboration reports that their current $B\to K\pi$ measurements are compatible with the Standard Model prediction for this rule \cite{BELLEKPI}, but the Belle II data set will improve the input measurements, particularly $\mathcal{A}_{CP}(B^{0}\to K^{0}_{S}\pi^{0})$, such that deviations of this sum rule can be probed at the few percent level.

\Acknowledgements
Firstly, I would like to thank Alexander Lenz who was a super co-convener. Secondly, all the speakers in WG5 - Guido Bell, Martin Sevior, Eli Ben-Haim, Marc Grabalosa, Jochim Brod, Aparajit Soni, Matteo Rama, Karim Trabelsi, Roy Briere, Moritz Karbach, Pablo Goldenzweig and Jaap Panman - for their excellent talks and the enjoyable discussions I had with many of them. Finally, I wish to thank Christoph Schwanda and the rest of the organizing  committees for the invitation to participate in this scientifically excellent and sociable workshop.

.


\begin{thebibliography}{99}
\bibitem{CC} Charge-congjugate modes are implied unless explicitly stated otherwise.
\end{thebibliography}
\end{document}